\documentclass[10pt,twocolumn,letterpaper]{article}

\usepackage{iccv}
\usepackage{times}
\usepackage{epsfig}
\usepackage{graphicx}
\usepackage{amsmath}
\usepackage{amssymb}

\usepackage{algorithm}
\usepackage[noend]{algorithmic}
\usepackage{booktabs}
\usepackage{array}
\usepackage{ragged2e}
\usepackage{caption}
\usepackage{subcaption}
\usepackage{wrapfig}
\usepackage[export]{adjustbox}

\usepackage[pagebackref=true,breaklinks=true,letterpaper=true,colorlinks,bookmarks=false]{hyperref}

\iccvfinalcopy % *** Uncomment this line for the final submission

\def\iccvPaperID{3529} % *** Enter the ICCV Paper ID here

\ificcvfinal\fi

\begin{document}

%%%%%%%%% TITLE
\title{Auto-FedAvg: Learnable Federated Averaging for Multi-Institutional Medical Image Segmentation}

\author{Yingda Xia\textsuperscript{1}\thanks{Work done during an internship at Nvidia},
Dong Yang\textsuperscript{2},
Wenqi Li\textsuperscript{2},
Andriy Myronenko\textsuperscript{2},
Daguang Xu\textsuperscript{2},
Hirofumi Obinata\textsuperscript{3},\\
Hitoshi Mori\textsuperscript{3},
Peng An\textsuperscript{4},
Stephanie Harmon\textsuperscript{5},
Evrim Turkbey\textsuperscript{6},
Baris Turkbey\textsuperscript{5},
Bradford Wood\textsuperscript{6},\\
Francesca Patella\textsuperscript{7},
Elvira Stellato\textsuperscript{8},
Gianpaolo Carrafiello\textsuperscript{8},
Anna Ierardi\textsuperscript{9},
Alan Yuille\textsuperscript{1},
Holger Roth\textsuperscript{2}
\vspace{0.2cm}
\\
\textsuperscript{1}Johns Hopkins University\quad\textsuperscript{2}NVIDIA\\ \quad\textsuperscript{3}Japan Self-Defense Forces Central Hospital\quad\textsuperscript{4}Xiangyang NO.1 People Hospital\\ \quad\textsuperscript{5}National Cancer Institute\quad\textsuperscript{6}National Institutes of Health \quad\textsuperscript{7}ASST Santi Paolo e Carlo \\\quad\textsuperscript{8}University of Milan\quad\textsuperscript{9}Foundation IRCCS Ospedale Maggiore Policlinico Hospital\\
}

\maketitle
\ificcvfinal\thispagestyle{empty}\fi

%%%%%%%%% ABSTRACT
\begin{abstract}
   Federated learning (FL) enables collaborative model training while preserving each participant's privacy, which is particularly beneficial to the medical field. FedAvg is a standard algorithm that uses fixed weights, often originating from the dataset sizes at each client, to aggregate the distributed learned models on a server during the FL process. However, non-identical data distribution across clients, known as the non-i.i.d problem in FL, could make this assumption for setting fixed aggregation weights sub-optimal. In this work, we design a new data-driven approach, namely~\textbf{Auto-FedAvg}, where aggregation weights are dynamically adjusted, depending on data distributions across data silos and the current training progress of the models. We disentangle the parameter set into two parts, local model parameters and global aggregation parameters, and update them iteratively with a communication-efficient algorithm. We first show the validity of our approach by outperforming state-of-the-art FL methods for image recognition on a heterogeneous data split of CIFAR-10. Furthermore, we demonstrate our algorithm's effectiveness on two multi-institutional medical image analysis tasks, i.e., COVID-19 lesion segmentation in chest CT and pancreas segmentation in abdominal CT.

\end{abstract}

%%%%%%%%% BODY TEXT
\section{Introduction}
Federated Learning (FL) \cite{mcmahan2017communication,yang2019federated,li2020federated} is a machine learning paradigm where clients collaboratively train a model without exchanging the underlying raw data. Compared to traditional centralized training, FL aims to benefit each participant while mitigating the potential for violating data privacy. FL was initially designed for mobile and edge devices~\cite{mcmahan2017communication} involving thousands of clients with often interrupted connectivity and only relatively small data each. However, recent studies involving only a small number of relatively reliable clients, e.g., medical institutions, have raised interest in utilizing FL for healthcare applications ~\cite{rieke2020future}. The latter scenario is referred to as ``cross-silo" FL in Kairouz et al.~\cite{kairouz2019advances} and is the focus of this paper.

\begin{figure}[!t]
\begin{center}
    \includegraphics[width=\columnwidth]{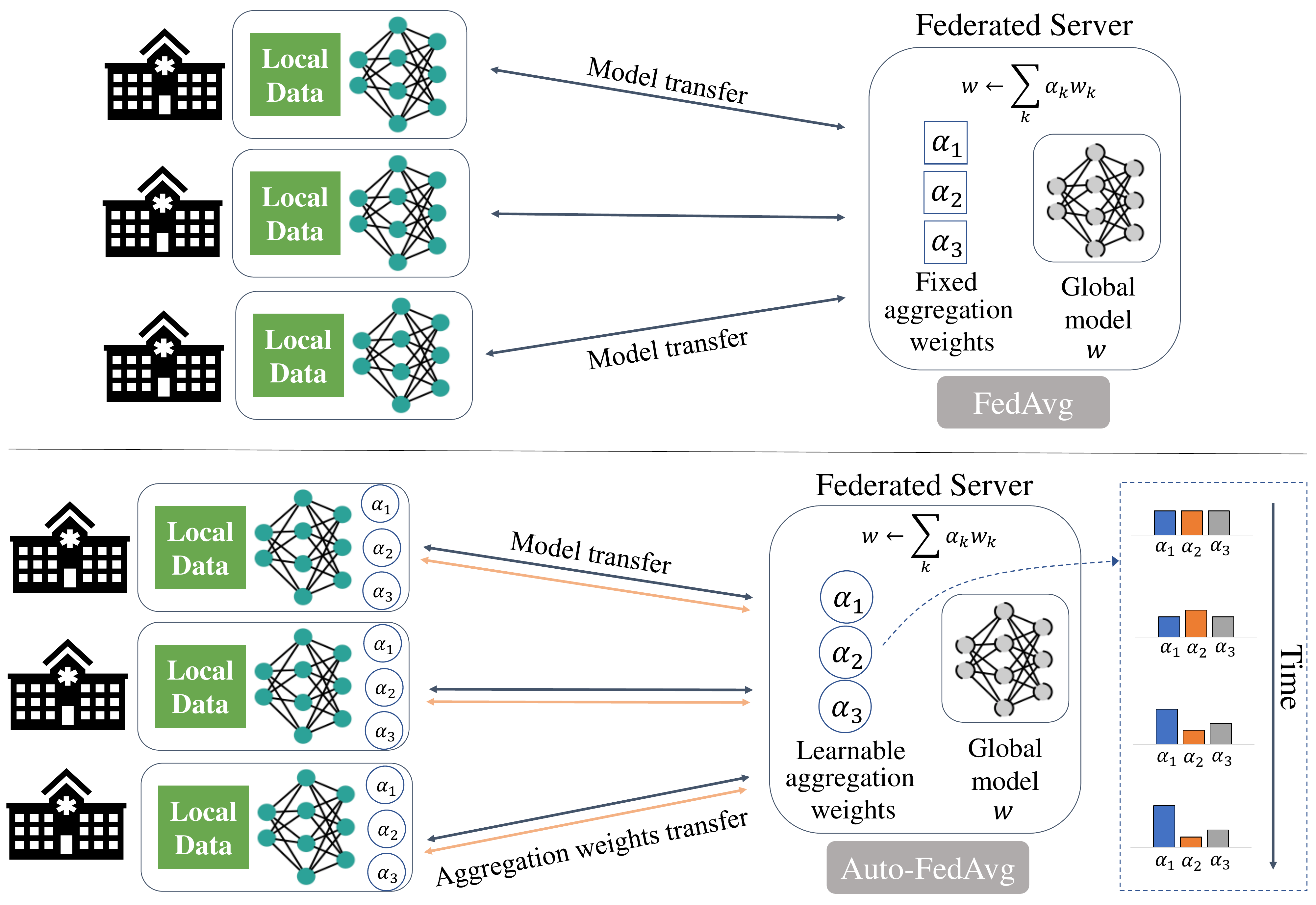}
\end{center}
\vspace{-0.2cm}
\caption{
    An illustration of FedAvg (top) and Auto-FedAvg (bottom). In FedAvg, the server collects locally trained models from each client and obtains a global model by weighted averaging with fixed aggregation weights. In contrast, in Auto-FedAvg, the aggregation weights are learned on the clients and dynamically adjusted throughout the training process when communicating with the server.
}
\vspace{-0.2cm}
\label{Fig:mo}
\end{figure}

Federated averaging (FedAvg)~\cite{mcmahan2017communication} is a simple yet effective algorithm for federated learning, following a server-client setup with two repeated stages: (i) the clients train their models locally on their data, and (ii) the server collects and aggregates the models to obtain a global model by weighted averaging. The aggregation weight of FedAvg is usually determined by the number of data samples on each client. This design choice assumes that data is uniformly distributed on the clients, and a stochastic gradient descent (SGD) optimizer is enforced. However, this setting can hardly be optimal and even detrimental because the clients' underlying data distributions remain unknown and are most likely non-independent and identically distributed (non-i.i.d). Domain shifts in the data are expected among different clients in real-world scenarios.

In this paper, we aim to improve FedAvg by automatically learning how to aggregate different client models more optimally. Our approach, namely Auto-FedAvg, is data-driven and differentiable while keeping the privacy-preserving aspects of FL. Recall that FedAvg involves two iterative steps. Our approach introduces a third step. After the clients finish training their local models, we learn a set of global aggregation weights in a data driven fashion, which the server later uses in the weighted average for computing the global model. Learning the global aggregation weights is beneficial in two aspects: (i) Since the convergence rate is likely to be different across the clients, dynamically adjusting aggregation weights can accelerate the training process. (ii) Better performance and generalizability can be achieved because the global model is more robust when applied to all the client's test data since we directly optimize the local loss to update the aggregation weights by modelling them as a stochastic process utilizing the Dirichlet distribution. We also designed a communication-efficient algorithm to achieve this goal without violating the data privacy constraint of FL.

%We show the effectiveness of our approach on two medical image segmentation tasks, \emph{i.e.}, multi-institutional and multi-national COVID-19 lesion segmentation and pancreas segmentation, where we outperform the FedAvg algorithm. We also find that the learned aggregation weights are interpretable in our analysis experiments. Moreover, we outperform the state-of-the-art method, FedMA~\cite{fedma}, on the CIFAR-10 dataset by 1.45\% using the same heterogeneous data partitioning.
We first validate the effectiveness of our approach on the CIFAR-10 dataset, where we outperform the state-of-the-art method, FedMA~\cite{fedma} by 1.45\% using the same heterogeneous data partitioning. Moreover, we outperform the FedAvg algorithm on two medical image segmentation tasks, \emph{i.e.}, multi-institutional and multi-national COVID-19 lesion segmentation and pancreas segmentation, showing its real-world potential.

Our contributions are summarized as follows:
\begin{itemize}
    \item We propose to directly learn the model aggregation weights in FL from data with gradient descent using a Dirichlet distribution, which is adaptive to the underlying data and learning progress.
    \item We design a new communication algorithm to fulfill the proposed goal with limited extra communication cost in cross-silo FL and without violating the data privacy constraints of FL.
    \item We outperform state-of-the-art approaches on a heterogeneous data split of CIFAR-10. Furthermore, we extensively analyze the proposed algorithm on two multi-institutional medical imaging studies with real-world datasets. %We also outperform state-of-the-art approaches on a heterogeneous data split of CIFAR-10. 
\end{itemize}

\section{Related Work}

%\noindent
%\textbf{Federated Learning} (FL) is defined in~\cite{kairouz2019advances} as a machine learning paradigm where multiple clients collaboratively train a model, typically under the orchestration of a central server. Data are stored in local devices or silos (e.g. data centers) and not exchanged or transferred to preserve privacy. FL was first introduced for use on mobile and edge devices~\cite{mcmahan2017communication} which could have unreliable connectivity and has been widely applied to multiple mobile applications~\cite{fedapp1, fedapp2}. Recently, the application of FL with a small number of relatively reliable clients (organizations) has raised increasing interests, such as medical image segmentation~\cite{fedmed1, fedmed2, fedmed3} and electronic health records mining~\cite{fedsilo1}. Based on the properties of participating clients, FL can be categorized into "cross-device" and "cross-silo" settings~\cite{kairouz2019advances}. In this paper, the application of our algorithm is targeted at "cross-silo" FL, where the clients are reliable and the communication bandwidth is less of a constraint.

\noindent
\textbf{Federated Learning.} 
Here, we introduce some common algorithms for FL. Federated Averaging (FedAvg)~\cite{mcmahan2017communication} is a standard algorithm, where parameters of local models are averaged with fixed weights to obtain a global model. The aggregation weight of each client is usually set to be proportional to the size of client's dataset. FedMA~\cite{fedma} refined the aggregation process by matching and averaging hidden elements with similar feature signatures. The idea of integrating knowledge distillation into FL has also been explored~\cite{fedkd1,fedkd2}.

Recently, the issue of FL on non-i.i.d data draws emerging attentions. Several works have been proposed to address data heterogeneity in FL settings~\cite{fednoniid1, fednoniid2, fednoniid3,Karimireddy2020,Chen2020}, among which one direction is to optimize the process of model aggregation that we also consider in this paper. For example, Wang et al.~\cite{Wang2020} proposed a normalized averaging method that eliminates objective inconsistency while preserving fast convergence for heterogeneous data clients. Chen et al.~\cite{Chen2020} analyzed median-based FL algorithms. Agnostic Federated Learning~\cite{AFL} proposed to optimize a centralized model for any target distribution formed by a mixture of the client distributions. FedBE~\cite{chen2020fedbe} learns a Bayesian ensemble from the distribution of the models. These works explore statistics or underlying distribution of the models to adjust aggregation strategies. In contrast, we propose to directly learn the aggregation weights by gradient-based optimization on the clients' data.
%Karimireddy et al.~\cite{Karimireddy2020} proposed to use a variance reduction technique to correct for the client-drift in its local updates.  Hanzely et al.~\cite{Hanzely2020} derived several optimal algorithms and their bounds for personalized FL. Dinh et al.~\cite{Dinh2020} proposed a personalized FL method using Moreau envelopes to improve global model performance in the presence of statistical diversity among clients. Wang et al.~\cite{Wang2020} proposed a normalized averaging method that eliminates objective inconsistency while preserving fast error convergence for heterogeneous data clients. Chen et al.~\cite{Chen2020} analyzed median-based FL algorithms, and provided a gradient correction mechanism that perturbs the local gradients to allow convergence on heterogeneous data.  FedProx~\cite{fedprox} improves FedAvg by adding a proximal term to regularize the training process of local models to be close to the global model. Agnostic Federated Learning~\cite{AFL} proposed to optimize a centralized model for any target distribution formed by a mixture of the client distributions. %APFL~\cite{APFL} aim a personalized model for each user that is a mixture of optimal local and global models.
Other recent works also discuss the possibility for model personalization~\cite{Hanzely2020, Dinh2020, APFL}.
Most recent works demonstrate good theoretical analysis but are only evaluated on manually created toy examples. It is not clear if the approaches would generalize well to real-world medical imaging datasets such as those studied in this work.
%In contrast, our approach aims to learn the aggregation weights of each client directly from data by gradient descent in a and communication-efficient way using real-world medical imaging datasets. 

\vspace{0.1cm}
\noindent
\textbf{Multi-institutional Medical Image Analysis.} 
Due to its privacy-preserving attributes, FL is particularly attractive for the medical domain. Rieke et al.~\cite{rieke2020future} discussed the potential of FL in digital health. Meanwhile, multiple real-world investigations of FL have been applied to medical image analysis, which is itself a well-explored field with deep learning~\cite{unet,v-net,3dunet}. Examples of FL in medical imaging include multi-institutional brain tumor segmentation~\cite{fedmed1,fedmed3}, breast density classification~\cite{fedmed2} and fMRI analysis~\cite{li2020multi}. In addition to FL settings, Chang et al.~\cite{privacygan} synthesized medical images with a GAN~\cite{gan} without sharing data between institutions. On top of privacy concerns, Liu et al.~\cite{liu2020ms}, Dou et al.~\cite{dou2019domain} and Xia et al.~\cite{xia2020uncertainty} emphasized the challenge of domain shift for multi-institutional medical data and developed algorithms to solve domain adaptation and generalization problems in prostate segmentation, brain tissue segmentation and liver segmentation from multi-site medical images, respectively. However, these non-i.i.d. challenges have not been resolved in FL for medical imaging \cite{rieke2020future}. 

\vspace{0.1cm}
\noindent
\textbf{Automated Machine Learning.}
This paper introduces an automated approach to find the best aggregation weights for federated learning. Our approach is inspired by recent advances of automated machine learning (AutoML), including hyper-parameter search~\cite{learningtolearn, autoaugment, randaugment}, neural architecture search (NAS) with reinforcement learning~\cite{nasrl1, nasrl2}, evolution algorithm~\cite{nasevo1, nasevo2} and differentiable approaches~\cite{nasdif1, nasdif2}. A recent approach~\cite{chen2020drnas} improves NAS by modeling the architecture mixing weight using a Dirichlet distribution, a mathematical formulation that we also utilize in this work. In the broad sense of AutoML, our approach can also be categorized as a differentiable hyper-parameter search algorithm in the continuous search space of FedAvg aggregation weights.

\section{Auto-FedAvg}
In this section, we first describe the general notations of federated learning and revisit FedAvg~\cite{mcmahan2017communication} in Sec~\ref{Sec:3.1}. We then introduce our optimization objective in Sec~\ref{Sec:3.2}, where we will also introduce how we parameterize the aggregation weights to follow certain constraints, as well as variants of the aggregation strategies, \emph{i.e.}, network-wise and layer-wise. Finally, in Sec~\ref{Sec:3.3}, we describe our full algorithm in detail and analyze the communication cost of the proposed Auto-FedAvg approach.

\subsection{Revisiting FedAvg}
\label{Sec:3.1}
Suppose $K$ clients collaboratively train a global model with parameter $w$ in a standard FL setting. In particular, the aim is to minimize:

\vspace{-0.3cm}
\begin{equation}
    \min_w\sum_{k=1}^K \alpha_k\mathcal{L}_k(w),
\end{equation}

where $\mathcal{L}_k(w)$ is the local loss function of client $k$, $\alpha_k \geq 0$ and $\sum_k \alpha_k=1$. Suppose there are $n_k$ data samples on client $k$, then we usually set $\alpha_k = \frac{n_k}{n}$, where $n=\sum_k n_k$ is the total number of data samples used in the FL setting.

To relieve the communication burden, FedAvg~\cite{mcmahan2017communication} allows the clients to update their local models for a certain period of time with the stochastic gradient descent (SGD) optimizer. We denote the local loss function given a data sample $x$ and the current model weight $w$ as $l(w, x)$.
The server then collects $C$ models ($C \leq K$), aggregates them with weighted averaging to update the global model, and sends the new global model back to the clients for re-initialization of next round of FL training. The aggregation weights $\alpha \in \mathbb{R}^{K}$ are set to be proportional to the number of data samples on each client ($\alpha_k = \frac{n_k}{n}$) as mentioned before. We pick $C=K$ for simplicity and the update of the global model $w$ in each communication round
%at communication round $t$ can be denoted 
as $w \leftarrow \sum_k \frac{n_k}{n} w_k$ where $w_k$ is the current model of client $k$.

The aggregation weights chosen by vanilla FedAvg is based on the assumption that data follows a uniform distribution across clients and are computed based on the number of SGD steps performed on each client. However, since the data distribution at each client is unknown and could possibly be non-i.i.d or involve domain shifts, this assumption is not guaranteed and can result in sub-optimal or even detrimental effects.

\subsection{Optimization Objectives}
\label{Sec:3.2}
To counteract the limitations of FedAvg, we propose our differentiable approach to directly learn the aggregation weights $\alpha$ from data at the clients. %Denote by $\mathcal{L}^{val}$ and $\mathcal{L}^{train}$ the validation and training loss. 
Denote by $\mathcal{L}$ the loss function. We propose a constrained objective function in:%on the distributed validation sets in:

\vspace{-0.5cm}
\begin{align}
    %\min_\alpha \indent& \mathcal{L}^{val}(\sum_{k=1}^{K} \alpha_k w_k) = \sum_{k=1}^K  \mathcal{L}^{val}_k(\sum_{k=1}^{K} \alpha_k w_k) \\
    \nonumber
    %\min_\alpha \indent& \sum_{k=1}^K  \mathcal{L}^{val}_k(\sum_{k=1}^{K} \alpha_k w_k) \\ 
    \min_\alpha \indent& \sum_{k=1}^K  \mathcal{L}_k(\sum_{k=1}^{K} \alpha_k w_k) \\ 
    \text{s. t.} \indent & \sum_{k=1}^K \alpha_k = 1 \ \mathrm{and} \ \alpha_k > 0,
\label{eq2}
\end{align}
%where $w_k = \underset{w}{\arg \min} \mathcal{L}^{train}_k(w)$ 
where $w_k = \underset{w}{\arg \min} \mathcal{L}_k(w)$
is the local model updated on the training set of client $k$.
%and $\mathcal{L}^{val}_k$ is the loss function on the validation set of client $k$. 
The motivation of the proposed objective is that we directly learn the aggregation weight by gradient descent from data in a differentiable way, while keeping the local models fixed after completing their local training. Since there is no data sharing between clients, we will introduce a communication algorithm to achieve the learning objective later in Sec~\ref{Sec:3.3}. We first discuss the variants of the constraints of Eq.~\ref{eq2} as follows.
%This is a constrained optimization problem and we introduce the variants of the constraints as follows.

%Optimize this objective function is not trivial under federated learning setting, since (i) we could only rely on the validation data on each client which is inaccessible to the server, and (ii) we should still maintain relatively low communication cost.

\vspace{-0.3cm}
\paragraph{Constraints of the aggregation weights.}
Here, we provide two assumptions for the optimization constraints in Eq.~\ref{eq2}. To achieve these constraints, we introduce a new set of variable $\pmb{\beta} = [\beta_1,..,\beta_K]$, which is a vector with the same dimension as $\pmb{\alpha} = [\alpha_1,..,\alpha_K]$. We define a function $\gamma$ to transform $\pmb{\beta}$ to $\pmb{\alpha}$:

\vspace{-0.2cm}
\begin{equation}
\label{eq:alpha}
    \pmb{\alpha} = \gamma(\pmb{\beta})
\end{equation}

\noindent\underline{Softmax function.} One obvious choice to satisfy the constraint of $\pmb{\alpha}$ is to apply a \textit{softmax} function to $\pmb{\beta}$
\begin{equation}
    \alpha_k = \frac{\exp{(\beta_k})}{\sum_{i=1}^K \exp{(\beta_i)}}
\label{eq:sm}
\end{equation}

Thus, the loss function becomes 
%$l(\sum_{k=1}^K \alpha_k w_k, x) = l(\sum_{k=1}^K \frac{\exp{(\beta_k})}{\sum_{k=1}^K \exp{(\beta_k)}} w_k, x) = \mathcal{L}(\pmb{\beta}, x)$,
$l(\sum_{k=1}^K \alpha_k w_k, x) = \mathcal{L}(\pmb{\beta}, x)$,
which only depends on $\beta$ and $x$, since we keep the model weights $w_k$ fixed in the aggregation weight learning process (Eq.~\ref{eq2}). In practice, we can compute the gradient of each $\beta_k$ and directly update them based on a client's local 
%validation 
data with gradient descent. 
%However, we will experimentally show that \textit{softmax} is an suboptimal choice to model the aggregation weights. 

\vspace{0.2cm}
\noindent\underline{Dirichlet distribution.} 
A better choice is to treat the aggregation weight $\pmb{\alpha}$ as random variables, modeled by the Dirichlet distribution parameterized by the concentration $\pmb{\beta}$: $\pmb{\alpha} \sim \mbox{Dir}(\pmb{\beta})$. This formulation induces stochasticity that naturally encourages exploration in the search space during the sampling process in training.
%A better choice might be to assume that $\pmb{\alpha}$ follows the Dirichlet distribution parameterized by the concentration $\pmb{\beta}$: $\pmb{\alpha} \sim \mbox{Dir}(\pmb{\beta})$. 
The probability density function is formed as:

\begin{equation}
    \mbox{Dir}(\pmb{\alpha} | \pmb{\beta}) = \frac{1}{B(\pmb{\beta})} \prod_{k=1}^K \alpha_k^{\beta_k - 1},
\end{equation}
where $B(\pmb{\beta}) = \frac{\prod_{k=1}^K \Gamma (\beta_k)}{\Gamma (\sum_{k=1}^K \beta_k)}$ and $\Gamma (z) = \int_0^\infty x^{z-1}e^{-x}dx $ is the gamma function. The Dirichlet distribution is the conjugate prior of a multinomial distribution with a simplex. Each sample will already satisfy our constraint of the aggregation weights in Eq.~\ref{eq2}. Thus we find the Dirichlet distribution to be a natural formulation to model the aggregation weights during FL while utilizing its properties for gradient-based optimization \cite{chen2020drnas,jankowiak2018pathwise}. It is also worth mentioning that the uniform distribution is a special case of the Dirichlet distribution when $\alpha_1=\alpha_2=...=\alpha_K=1$.

In the training phase, given a data sample $x$, we sample $\pmb{\alpha}$ from the Dirichlet distribution with concentration $\pmb{\beta}$, approximate the gradient of $\pmb{\beta}$ given the loss function $\mathcal{L}(\pmb{\beta}, x)$ using implicit reparameterization~\cite{figurnov2018implicit}
 and update the concentration $\pmb{\beta}$. During inference, we compute the mode of the distribution, which represents the values with maximum probability. 

\begin{equation}
    \alpha_k = \frac{\beta_k - 1}{\sum_{i=1}^K \beta_i - K}
\label{eq:dirinfer}
\end{equation}

\vspace{-0.5cm}
\paragraph{Aggregation strategies.} In the process of model aggregation, our approach introduces more flexibility in terms of the design of the aggregation weights than FedAvg, because we are able to learn the parameterized aggregation weights in a differentiable way from data. Here, we describe two natural variants.

\vspace{0.1cm}
\noindent\underline{Network-wise aggregation weights.} In this scenario, each aggregation weight $\alpha_k$ in $\pmb{\alpha}$ is a scalar. The aggregation process is the same as described previously: $w \leftarrow \sum_k \alpha_k w_k$.

\vspace{0.1cm}
\noindent\underline{Layer-wise aggregation weights.} Our approach allows an easy extension to network-wise aggregation, namely layer-wise aggregation. Suppose the deep network model we are training has $P$ layers. We denote $w_{k,p}$ as the $p$-th layer parameter of the model of client $k$. Then $\alpha_k = [\alpha_{k,1},..,\alpha_{k,P}]$ is a $P$-dimensional vector. Thus we are able to obtain the $p$-th layer weight $w_p$ by $w_p \leftarrow \sum_{k=1}^K \alpha_{k,p} w_{k,p}$. 

As for the constraints discussed previously, $\beta_k = [\beta_{k,1},..,\beta_{k,P}]$ is now a $P$-dimensional vector as well. Then, $\alpha_{k,p} = \frac{\exp{(\beta_{k,p}})}{\sum_{k=1}^K \exp{(\beta_{k,p})}}$ is the equation when using softmax, and $\pmb{\alpha}_p \sim \mbox{Dir}(\pmb{\beta}_p)$ when applying the Dirichlet distribution.

\begin{algorithm}[t]
\caption{Auto-FedAvg. We denote the total number of rounds as $T$, the interval to learn aggregation weights as $t_0$, local training iterations for client $k$ as $M_k$, and the aggregation weight learning iterations as $S$.}
\textbf{Server executes:}
\begin{algorithmic}
\STATE Define $\pmb{\alpha}^t = [\alpha_1^t,..,\alpha_K^t]$, $\pmb{\beta}^t = [\beta_1^t,..,\beta_K^t]$. \\
Initialize $w^0$ and $\pmb{\beta}^0$. $\pmb{\alpha}^0 = \gamma (\pmb{\beta}^0)$ \\
\FOR{$t \leftarrow 1,...,T$} 
\FOR{$k \leftarrow 1,...,K$ \textbf{in parallel}}
%\STATE Send $w^{t-1}$ to client $k$
\STATE $w_k^t \leftarrow$ \texttt{LocalTrain}$(k, w^{t-1})$
\ENDFOR
\IF{$t \bmod t_0 = 0$}
\STATE $\pmb{\beta}^t \leftarrow$ \texttt{LearnAggWeight}$(w_1^t,..,w_K^t, \pmb{\beta}^{t-1})$
\STATE $\pmb{\alpha}^t \leftarrow \gamma(\pmb{\beta}^t)$
\ELSE 
\STATE $\pmb{\alpha}^t \leftarrow \pmb{\alpha}^{t-1}$
\ENDIF
\STATE $w^t \leftarrow \sum_{k=1}^K \alpha_k^t w_k^t$
\ENDFOR
\RETURN $w^T$
\end{algorithmic}

\vspace{0.05cm}
\texttt{LocalTrain}($k, w$): 
\begin{algorithmic}
\FOR{$t\leftarrow 1,..,M_k$}
\STATE Sample batch $x$ from client $k$'s training data
\STATE Compute loss $l(w; x)$
\STATE Compute gradient of $w$ and update $w$
\ENDFOR
\RETURN $w$
\end{algorithmic}

\vspace{0.05cm}
\texttt{LearnAggWeight}($w_1,..,w_K, \pmb{\beta}^0$): 
\vspace{-0.4cm}
\begin{algorithmic}
\FOR{$k \leftarrow 1,...,K$}
\STATE Server send $w_1, .., w_{k-1}, w_{k+1},.., w_K$ to client $k$
\ENDFOR
\FOR{$s\leftarrow 1,..,S$}
\FOR{$k \leftarrow 1,...,K$  \textbf{in parallel}}
\STATE Sever send $\pmb{\beta}^{s-1}$ to client $k$
\STATE Sample batch $x$ from client $k$'s local data
\STATE Compute loss $\mathcal{L}(\pmb{\beta}^{s-1}; x)$
\STATE Compute/estimate gradient and update $\pmb{\beta}^{s-1}$ as $\pmb{\beta}^{s,k}$
\vspace{-0.4cm}
\STATE Send $\pmb{\beta}^{s,k}$ back to the server
\ENDFOR
$\pmb{\beta}^s \leftarrow \frac{1}{K}\sum_{k=1}^K \pmb{\beta}^{s,k}$
\ENDFOR
\vspace{0.05cm}
\RETURN $\pmb{\beta}^S$
\end{algorithmic}

\label{algo}
\end{algorithm}

\subsection{Algorithm}
\label{Sec:3.3}
Optimizing the objective function in Eq.~\ref{eq2} is not trivial under the FL setting, since (i) we can only rely on the local data on each client which is inaccessible to the server, and (ii) we would like to maintain a relatively low communication cost. We describe the algorithm of Auto-FedAvg in Algorithm~\ref{algo}. In each communication round $t$, the server first sends out the global model to all the clients. When the clients finish updating the local models in parallel, the server gathers them and aggregates the models with a set of learnable weights $\pmb{\alpha}^t = [\alpha_1^t,..,\alpha_K^t]$ by weighted averaging to obtain an updated global model $w^t$. $\pmb{\alpha}^t$ is parameterized by $\pmb{\beta}^t$ using function $\gamma$ and the actual instantiation of $\gamma$ in Eq.~\ref{eq:alpha} is determined by whether we use softmax (Eq.~\ref{eq:sm}) or the Dirichlet distribution (Eq.~\ref{eq:dirinfer}) as the method to parameterize $\alpha$. The learning process of $\pmb{\beta}^t$ is described in \texttt{LearnAggWeight} of Algorithm~\ref{algo}.

In \texttt{LearnAggWeight}, each client receives a copy of all the model weights $w_1, ..,w_K$ and keeps them fixed during this process. In each local iteration $s$, each client samples a mini-batch $x$ from their own local data, and computes the current $\pmb{\alpha}$ from $\pmb{\beta}^{s-1}$ depending on the softmax or Dirichlet assumption we apply to the aggregation weights, before forwarding $x$ into the local model with weight $\sum_{k=1}^{K} \alpha_k w_k$. Then the client will compute the loss function $\mathcal{L}(\pmb{\beta}^{s-1}, x)$ and update $\pmb{\beta}^{s,k}$ based on the computation (softmax) or estimation (Dirichlet distribution) of the gradient~\cite{figurnov2018implicit}, as mentioned in Sec~\ref{Sec:3.2}. The server will gather $\pmb{\beta}^{s,k}$ from every client $k$ in every iteration $s$ and average them to obtain a new global $\pmb{\beta}^s$. 

\vspace{-0.5cm}
\paragraph{Communication efficiency analysis.}
The communication of $\pmb{\beta}$ is very efficient because $\pmb{\beta}$ is merely a set of $K$ scalars or $K$ low dimensional vectors (of size $P$) in either ``network-wise aggregation weight" or ``layer-wise aggregation weight" strategy, which is negligible compared to communicating the full network parameters as in a standard FedAvg round. The major extra communication burden of aggregation weight learning is introduced when the server sends all local models to each client in the very first step. As a result, we only do the aggregation weight learning process every $t_0$ rounds to further relieve the additional communication burden compared to FedAvg. The extra communication cost ratio (extra cost divided by FedAvg communication cost) is $\frac{K-1}{2t_0}$. A detailed derivation of which can be found in the supplementary material. This is more acceptable in cross-silo federated learning setting, which typically contains only a small number of clients with relatively reliable internet connectivity \cite{kairouz2019advances}. For example, in our COVID-19 lesion segmentation experiments, $K=3$ and $t_0=10$, results in an extra 10\% communication cost compared to FedAvg.

\section{Experiments}

\subsection{CIFAR-10}
We first validate our approach on the CIFAR-10 dataset. To compare our approach with the state-of-the-art FL methods such as FedProx~\cite{fedprox} and FedMA~\cite{fedma} on the benchmark dataset, we use the same heterogeneous data partition of FedMA~\cite{fedma} on the CIFAR-10 dataset that simulates an environment where the number of data points and class proportions are unbalanced using their publicly available code\footnote{\url{https://github.com/IBM/FedMA}}. In this way, we can directly compare with the results in the paper, which are shown in Table~\ref{tab:CIFAR}. We train the baseline and our experiments for 99 rounds with 16 clients before we test on the test set, where the same network architecture of VGG-9 is adopted. The re-implementation of FedMA achieves 87.47\% accuracy, which is very close to the reported performance 87.53\% \cite{fedma}, indicating the correctness of our experimental setup. For our Auto-FedAvg algorithm, we experiment with different design choices described in Sec~\ref{Sec:3.2}, \emph{i.e.} layer-wise (``L") or network-wise (``N") aggregation strategy and softmax (``Softmax") or Dirichlet assumption (``Dirichlet") over the constraints of the aggregation weights. Based on the metric of final accuracy, all our experimental variants outperform the baselines and our ``Auto-FedAvg-N-Dirichlet" achieved the best final accuracy of 88.98\%, outperforming the published FedMA result by 1.45\%. 

\begin{table}[!t]
    \centering
    \caption{CIFAR-10 classification with heterogeneous partition. Baseline numbers are from \cite{fedma} on the same data split.}
    \vspace{-0.2cm}
\footnotesize
\begin{tabular}{l c c }
\hline
%\multicolumn{1}{l|}{} & \multicolumn{3}{c}{local test} &\\
%\cmidrule(lr){2-4}
\multicolumn{1}{l|}{Method}  &  final accuracy(\%) \\
\hline\hline
\multicolumn{1}{l|}{FedAvg} & 86.29  \\
\multicolumn{1}{l|}{FedProx~\cite{fedprox}} & 85.32 \\
\multicolumn{1}{l|}{Ensemble} & 75.29 \\
\multicolumn{1}{l|}{FedMA~\cite{fedma}} & 87.53 \\
\multicolumn{1}{l|}{FedMA~\cite{fedma} (our impl.)} & 87.47 \\
\hline\hline
\multicolumn{1}{l|}{Auto-FedAvg-L-Softmax*} &  88.64\\
\multicolumn{1}{l|}{Auto-FedAvg-L-Dirichlet*} &  88.37\\
\multicolumn{1}{l|}{Auto-FedAvg-N-Softmax*} &  88.60\\
\multicolumn{1}{l|}{Auto-FedAvg-N-Drichlet*} &  \textbf{88.98}\\
\hline

\end{tabular}
\\ $^*$ With the interval of aggregation weight learning $t_0 = 10$.
\vspace{-0.2cm}
\label{tab:CIFAR}
\end{table}

\newcommand\tabwidth{0.09\textwidth}
\newcommand{\RM}[1]
    {\MakeUppercase{\romannumeral #1}}
\newcolumntype{P}[1]{>{\centering\arraybackslash}p{#1}}
\begin{table*}[!t]
    \centering
    \caption{Multi-national COVID-19 lesion segmentation. ``Global test avg" is the major metric to measure the generalizability of the FL global model. $n$ specifies the total dataset size at the client.}
    \vspace{-0.2cm}
\footnotesize
    \begin{tabular}{l P{\tabwidth} P{\tabwidth} P{\tabwidth} P{0.16\textwidth} P{\tabwidth} P{\tabwidth}}
\hline
\multicolumn{1}{l|}{Method}  &  \RM{1} ($n$=671) & \RM{2} ($n$=88) & \RM{3} ($n$=186) & global test avg & local avg & local gen\\
\hline\hline
\multicolumn{1}{l|}{Local only - \RM{1}} & 59.82 &61.82 &51.80& 57.81\\
\multicolumn{1}{l|}{Local only - \RM{2}} & 41.92 &59.95 &50.18& 50.68& 61.87 &48.79\\
\multicolumn{1}{l|}{Local only - \RM{3}} & 34.50 &52.54 &\textbf{65.85}& 50.96\\
\hline\hline
\multicolumn{1}{l|}{FedAvg} & 59.93	&63.79	&60.52	&61.41 $\pm 0.19$&62.47	&58.80\\
\multicolumn{1}{l|}{FedAvg - even} & 56.73	&64.31	&64.98	&62.01 $\pm 0.30$ &62.24   &59.28\\
\multicolumn{1}{l|}{FedProx} & 60.33	&64.98	&60.45	&61.92 $\pm 0.53$&61.99	&58.33\\
\hline\hline
\multicolumn{1}{l|}{Auto-FedAvg-L-Softmax} & 59.03 &64.96$^\dagger$ &61.66$^\dagger$ &61.89 $\pm 0.54$&63.17	&58.96\\
\multicolumn{1}{l|}{Auto-FedAvg-L-Dirichlet} & 58.59 &64.95$^\dagger$	&64.96$^\dagger$	&62.83 $\pm 0.14$ &63.08 &59.51\\
\multicolumn{1}{l|}{Auto-FedAvg-N-Softmax} & 59.58 & 64.50$^\dagger$ & 63.35$^\dagger$ & 62.48 $\pm 0.24$ & 63.42 & 59.62\\
\multicolumn{1}{l|}{Auto-FedAvg-N-Dirichlet} & 60.37 &\textbf{65.28}$^\dagger$  &64.76$^\dagger$  &\textbf{63.47}$^\dagger$ $\pm 0.22$   &\textbf{64.04}   &\textbf{60.79}\\
\multicolumn{1}{l|}{Auto-FedAvg-N-Dirichlet*} & \textbf{60.42} &64.86$^\dagger$	&64.07$^\dagger$ &63.11$^\dagger$ $\pm 0.33$ & 63.74	&60.23\\
\hline

% FedAVG - 61.90
% FedAVG even - 62.85

\end{tabular}
\\ $^*$ With the interval of aggregation weight learning $t_0 = 10$. \\ $^{\dagger}$ Significance of the global model over FedAvg.
%\vspace{-0.2cm}
\label{tab:covid}
\end{table*}

\subsection{Multi-national COVID-19 lesion segmentation}
\label{Sec:covid}

\subsubsection{Experimental results}

The study with first real-world data of our federated learning algorithm is COVID-19 diagnosis, which has caused a world-wide pandemic in the year of 2020 and 2021. Machine learning based algorithms have been developed to quickly diagnose the disease and study the imaging characteristics \cite{harmon2020artificial,li2020artificial,shi2020review}. In this study, we focus on the critical task of COVID-19 lesion segmentation on multi-national COVID-19 datasets.

\begin{figure}[h]
\begin{center}
    \includegraphics[width=\columnwidth]{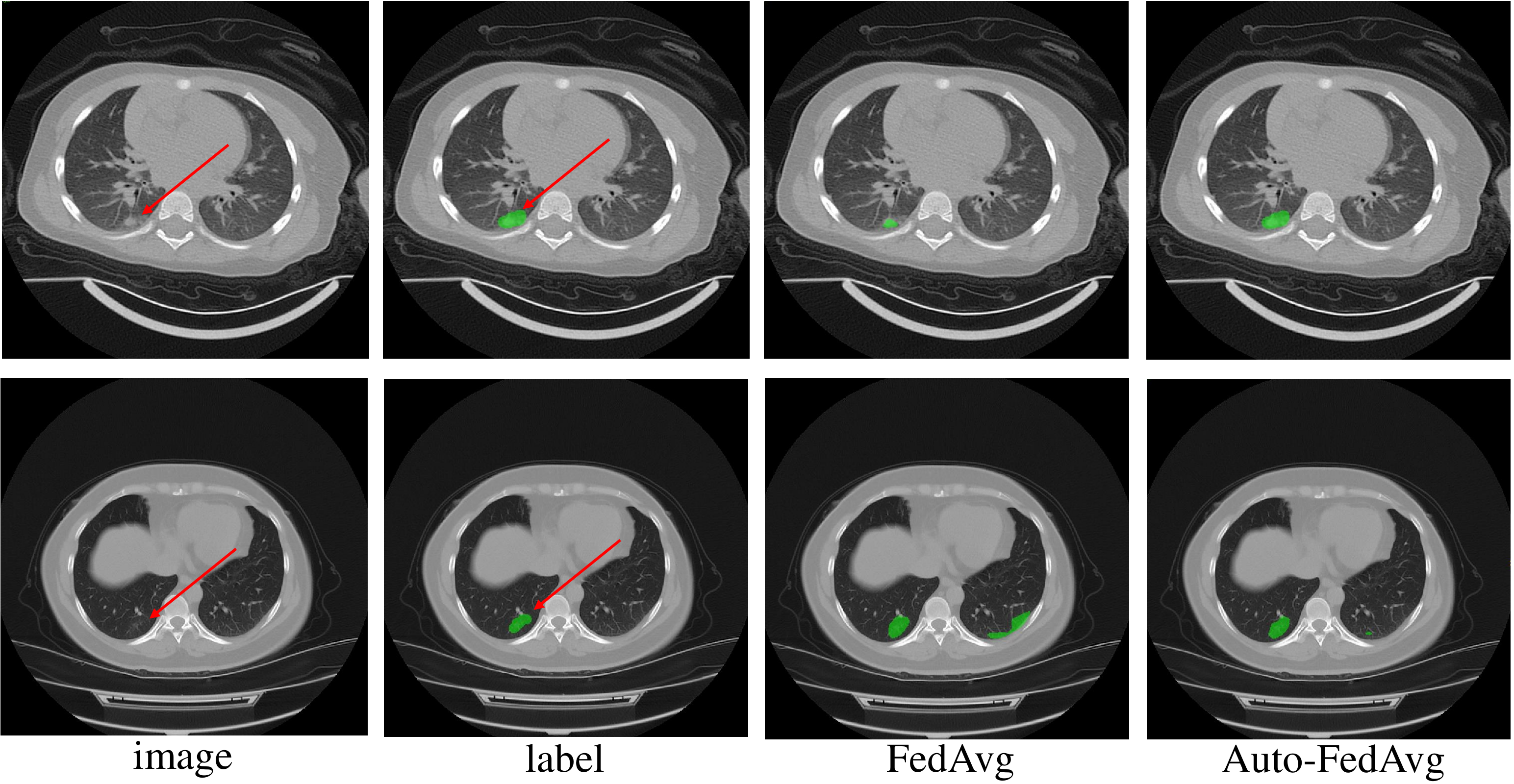}
\end{center}
\vspace{-0.5cm}
\caption{
    Examples of COVID-19 lesion segmentation of patients from China (top) and Italy (bottom). From left to right: original CT scan, human label (in green), FedAvg segmentation results, and our segmentation results. Our Auto-FedAvg mitigates the issue of under-segmentation (top) and reduces false-positive prediction (bottom) in these two examples, respectively.
}
\vspace{-0.5cm}
\label{Fig:covid}
\end{figure}

%\begin{figure*}[!t]
%\begin{center}
%    \includegraphics[width=8cm, height=4cm]{example-image-c}
%\end{center}
%\caption{
%    Figures show (i) the learning curve of the aggregation weights (ii) validation accuracy - rounds (iii) performance - interval.
%}
%\label{Fig:learn}
%\end{figure*}

\vspace{0.1cm}
\noindent\textbf{Dataset description.} This study contains CT scans of SARS-CoV-2 infected patients collected from three international medical centers, including 
%(i) the First Affiliated Hospital of Hubei University of Medicine in Hubei Province, China (denoted as Dataset \RM{1}), (ii) the Self-Defense Forces Central Hospital, Tokyo, Japan (denoted as Dataset \RM{2}), and (iii) San Paolo Hospital, Milan, Italy (denoted as Dataset \RM{3}). 
(i) 671 scans from [anonymized hospitals] in China (denoted as Dataset \RM{1}), (ii) 88 scans from [anonymized hospitals] in Japan (denoted as Dataset \RM{2}), and (iii) 186 scans from [anonymized  hospitals] in Italy (denoted as Dataset \RM{3}). 
Two expert radiologists annotated these CT scans assigning a foreground (COVID-19 lesion) and background label for each voxel. For each dataset, we randomly split the annotated cases into training/validation/testing, resulting in splits of 447/112/112 for Dataset \RM{1}, 30/29/29 for Dataset \RM{2}, and 124/31/31 for Dataset \RM{3}. We visualize examples in Fig~\ref{Fig:covid} and show the intrinsic domain shift between datasets (e.g., caused by resolution and contrast).

\vspace{0.1cm}
\noindent\textbf{Implementation details.} We simulate a federated learning scenario, in which each dataset represents one FL client, and ensure no data is transferred among the clients. In all FL experiments, we fix the number of total communication rounds $T$ to 300 and validate on the local validation sets in each round to select the best local models (with highest validation score on each single client) and the global model (with highest average validation accuracy over three validation sets). In our experiments, we initialize the concentration $\beta$ of the Dirichlet distribution as (6.0, 6.0, 6.0). In the local training process of each client, we adopt the Adam optimizer~\cite{kingma2014adam} with a learning rate of 0.0001, ($\beta_1, \beta_2$) as (0.5, 0.99), and no decay. Each training round performs 300 iterations with a batch size of 4. These hyperparameters are tuned to achieve the best local performances. Dice loss~\cite{v-net} is used as the training objective, which is a widely-applied loss function in medical image segmentation and aimed to handle the problem of foreground-background imbalance. The architecture of the segmentation network is 3D U-Net~\cite{unet,3dunet}. In the training process, we resample each CT volume to a fixed voxel spacing of (0.8mm, 0.8mm, 5mm) and randomly crop region of interests (ROIs) of $256 \times 256 \times 32$. In the testing phase, we adopt a sliding window scheme with a stride of (64, 64, 8) and resample to the original voxel spacing for final evaluation.

\vspace{0.1cm}
\noindent\textbf{Evaluation metrics.} We measure the performance of the segmentation models by Dice similarity coefficient (DSC), a standard evaluation metric used for medical image segmentation. For all the FL experiments, we test the performance of the best global model, selected by highest average validation accuracy of all three clients, on the test data of each client, corresponding to the first three columns (\RM{1}/\RM{2}/\RM{3}) of Table~\ref{tab:covid}. We compute the average of the three test accuracies to measure the average performance of the model on three datasets, corresponding to the forth column ``global test avg". This metric represents a measure for the generalizability of the global model, and serves as the major metric for performance evaluation. Moreover, we test the best local models on all clients, selected by the highest local validation score, resulting in a 3x3 matrix of scores. The on-diagonal scores represent the local performances of each model, which is averaged as ``local avg" in column five. The off-diagonal scores represent the generalization performance of each model, which is averaged as ``local gen" in column six. We run three repeats of each configuration of the FL experiments and report the standard deviation on ``global test avg" to measure the stability of our results.

\newcommand\figheight{3cm}
\begin{figure*}[!t]
     \centering
     \begin{subfigure}[b]{0.24\textwidth}
         \centering
         \includegraphics[height=\figheight]{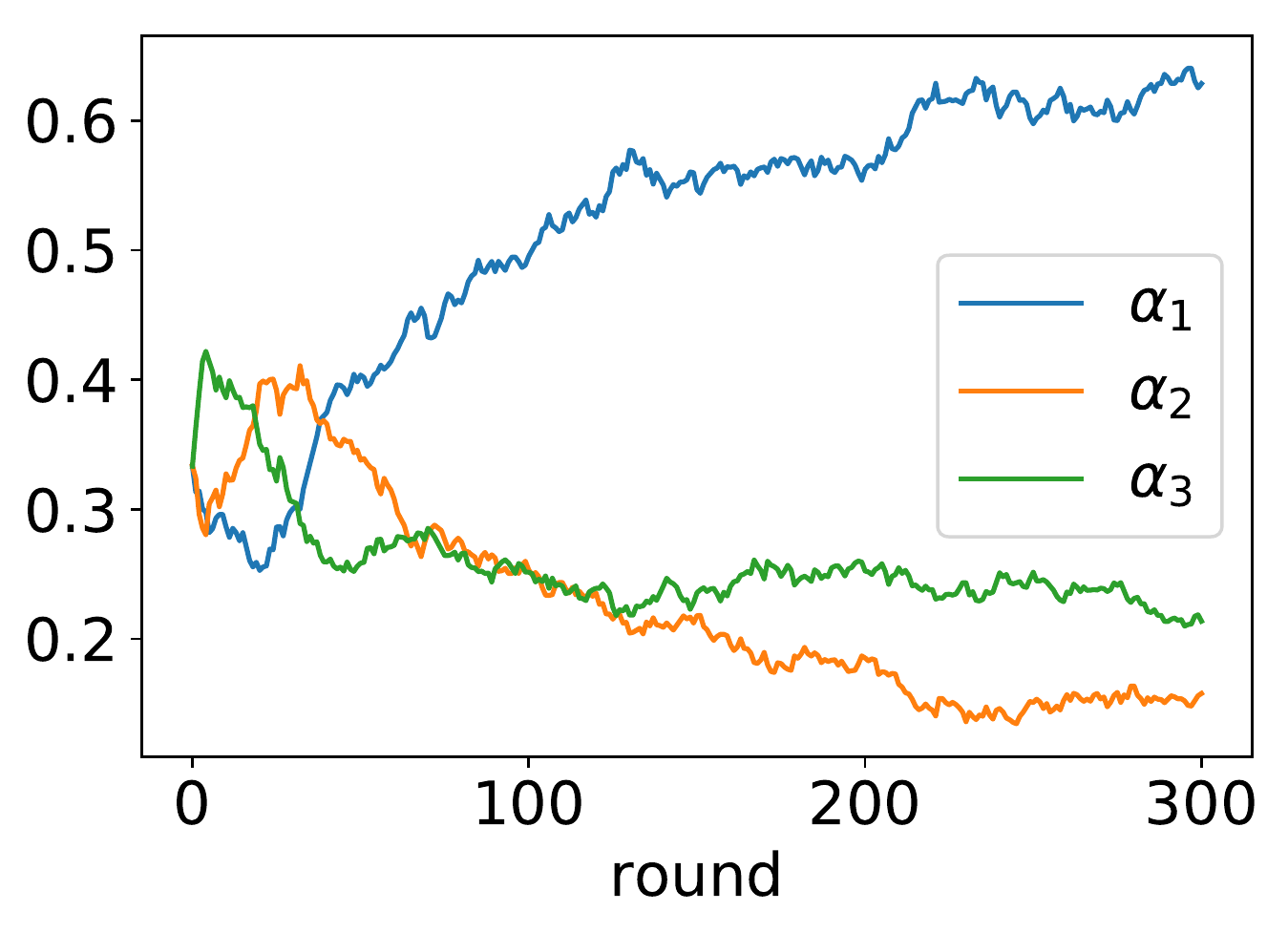}
         \caption{The learning curve of $\alpha$.}
         \label{fig:3a}
     \end{subfigure}
     \hfill
     \hspace{-0.25cm}
     \begin{subfigure}[b]{0.26\textwidth}
         \centering
         \includegraphics[height=\figheight]{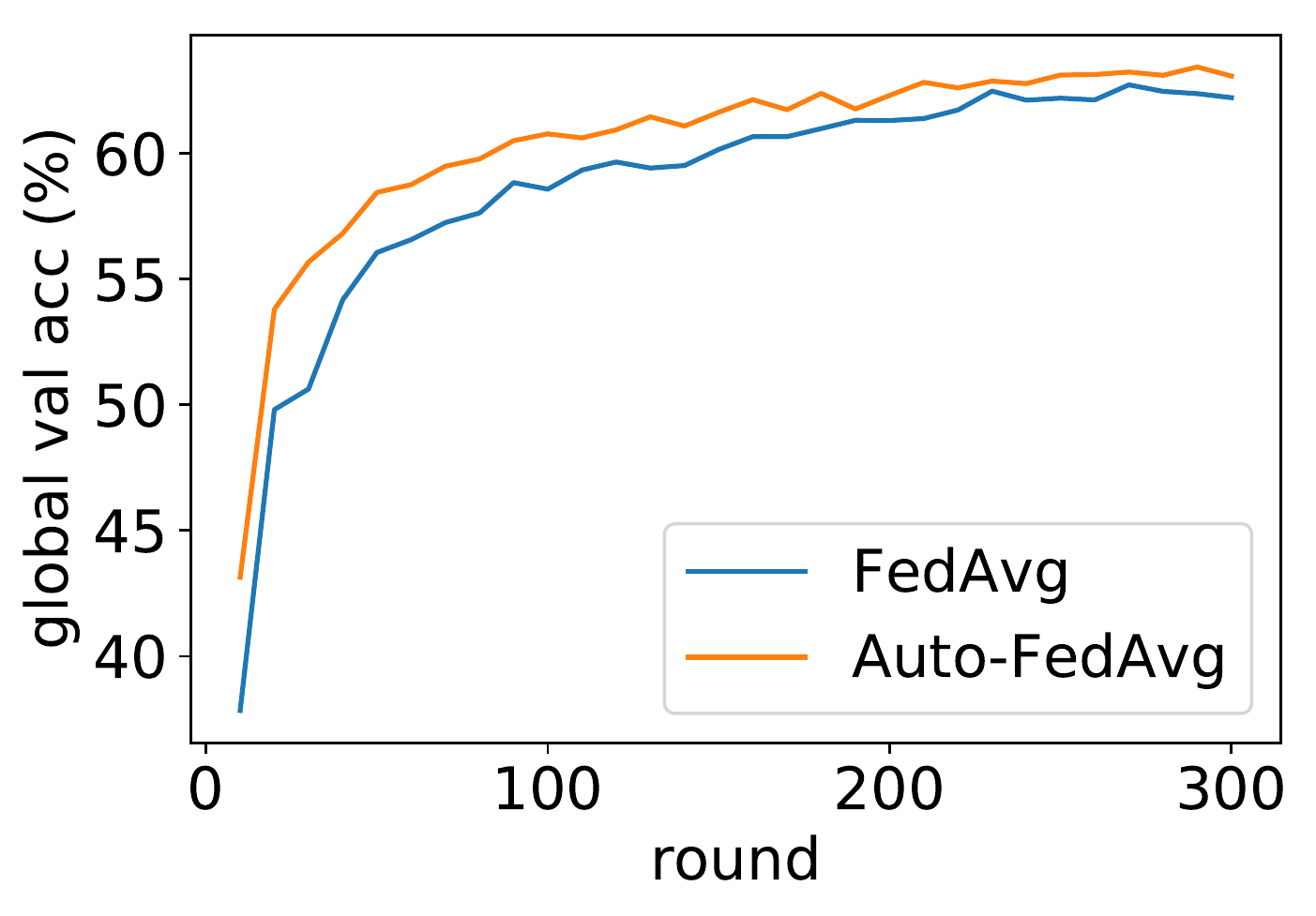}
         \caption{Validation accuracy growth.}
         \label{fig:3b}
     \end{subfigure}
     \hfill
     \begin{subfigure}[b]{0.49\textwidth}
         \centering
         \includegraphics[width=\columnwidth]{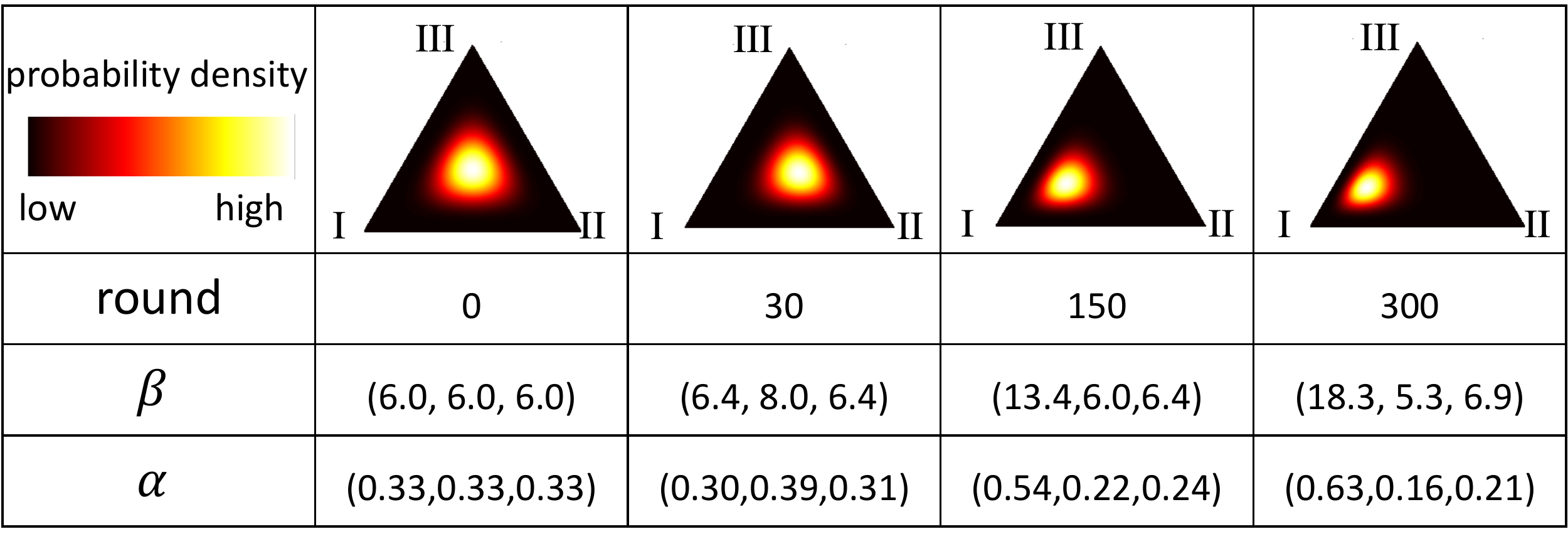}
         \caption{Visualizations of Dirichlet distribution.}
         \label{fig:3c}
     \end{subfigure}
\vspace{-0.2cm}
\caption{Analysis of the learning process during ``Auto-FedAvg-N-Dichlet".}
\vspace{-0.2cm}
\label{Fig:learn}
\end{figure*}

\vspace{0.1cm}
\noindent\textbf{Results.} We display the quantitative results in Table~\ref{tab:covid} and two examples for qualitative analysis in Fig~\ref{Fig:covid}. We first train the models locally without communication to obtain the baselines of the local models, shown in the first three rows in the table. Unsurprisingly, all three local models have relatively low generalization performance when tested on other clients, indicating domain shifts across the three datasets. For the FedAvg baseline, we experiment with two different sets of aggregation weights, \emph{i.e.}, normalized dataset size and uniform weights, denoted by ``FedAvg" and ``FedAvg-even", respectively. We also implement FedProx~\cite{fedprox} with the empirically best $\mu=0.001$. For our Auto-FedAvg algorithm, we experiment with different design choices described in Sec~\ref{Sec:3.2}, \emph{i.e.} layer-wise (``L") or network-wise (``N") aggregation strategy and softmax (``Softmax") or Dirichlet assumption (``Dirichlet") over the constraints of the aggregation weights. We find that ``Auto-FedAvg-N-Dirichlet" gives the best results, outperforming ``FedAvg" by 2.06\% on general global model performance (column "global test avg"), by 1.57\% on average local model performance (column ``local avg"), and by 1.99\% on local model generalization (column ``local gen"). We furthermore performed a Wilcoxon signed rank test on the test set (first four columns), where the significant improvements ($p \ll 0.05$) over FedAvg are marked with superscript $^\dagger$.

Generally speaking, the Dirichlet distribution performs better at modeling the aggregation weights than softmax. Interestingly, the performance of the layer-wise aggregation strategy is worse than the network-wise aggregation strategy. The gradient of network-wise aggregation weights can be viewed as a summation of all gradients of layer-wise aggregation weights. In this sense, we suspect that network-wise aggregation acts as a regularization of layer-wise weights. We also conduct diagnosis experiments and provided them in supplementary materials, where we display the patterns of the learned layer-wise weights and suggest a layer-wise smoothing loss can improve the results of the layer-wise aggregation strategy. The improvement of the layer-wise smoothing loss for the layer-wise aggregation strategy further serves as evidence that the network-wise aggregation may act as regularization over the layer-wise one.

%\begin{figure}[!t]
%\begin{center}
%    \includegraphics[width=\columnwidth]{figures/Fig4_v2.pdf}
%\end{center}
%\caption{
%    (a) The learning curve of $\alpha$, (b) validation accuracy growth and (c) visualizations of Dirichlet distribution during the training process.
%}
%\label{Fig:learn}
%\end{figure}

\begin{table*}[!t]
    \centering
    \caption{Multi-institutional pancreas segmentation. ``Global test avg" is the major metric to measure the generalizability of the FL global model. $n$ specifies the total dataset size at the client.}
    \label{tab:pancreas}
    \vspace{-0.2cm}
\footnotesize
    \begin{tabular}{l P{\tabwidth} P{\tabwidth} P{\tabwidth} P{0.16\textwidth} P{\tabwidth} P{\tabwidth}}
\hline
%\multicolumn{1}{l|}{} & \multicolumn{3}{c}{local test} &\\
%\cmidrule(lr){2-4}
\multicolumn{1}{l|}{Method}  &  \RM{1} ($n$=281) & \RM{2} ($n$=82) & \RM{3} ($n$=30) & global test avg & local avg & local gen\\
\hline\hline
\multicolumn{1}{l|}{Local only - \RM{1}} & 69.43	&71.38  &63.79 &  68.20  &   &\\
\multicolumn{1}{l|}{Local only - \RM{2}} & 49.69    &75.47	&53.02 &  59.39  &  65.32 & 56.9\\
\multicolumn{1}{l|}{Local only - \RM{3}} & 42.35	&61.18	&51.08 &  51.34  &   &\\
\hline\hline
\multicolumn{1}{l|}{FedAvg} & 71.85	&78.36	&69.12	&73.11 $\pm 0.17$	&72.84	&70.75\\
\multicolumn{1}{l|}{FedAvg - even} & 69.18	&78.82	&70.91	&72.97 $\pm 0.14$	&73.49	&71.33\\
\multicolumn{1}{l|}{FedProx} & \textbf{71.96}	&78.35	&69.57	&73.29 $\pm 0.25$&73.66	&70.79\\
\hline\hline
\multicolumn{1}{l|}{Auto-FedAvg-L-Softmax} & 71.22	&78.38	&71.04$^\dagger$	&73.54 $\pm 0.28$&73.92	&71.49\\
\multicolumn{1}{l|}{Auto-FedAvg-L-Dirichlet} & 71.06    &79.60$^\dagger$	&70.56$^\dagger$	&73.74 $\pm 0.34$	&74.17	&71.68\\
\multicolumn{1}{l|}{Auto-FedAvg-N-Softmax} & 70.40	&78.80	&70.48$^\dagger$	&73.22 $\pm 0.21$ &74.02	&71.50\\
\multicolumn{1}{l|}{Auto-FedAvg-N-Dirichlet} & 71.20	&79.30$^\dagger$	&71.19$^\dagger$	&73.90	$\pm 0.25$ &74.25	&71.83\\
\multicolumn{1}{l|}{Auto-FedAvg-N-Dirichlet*} & 71.26	& \textbf{79.90}$^\dagger$	&\textbf{71.49}$^\dagger$	&\textbf{74.21}	$\pm 0.28$&\textbf{74.33}	&\textbf{72.30}\\
\hline

\end{tabular}
\\ $^*$ With the interval of aggregation weight learning $t_0 = 5$. \\ $^{\dagger}$ Significance of the global model over FedAvg.
%\vspace{-0.3cm}
\end{table*}

\subsubsection{Analysis studies}
\vspace{-0.2cm}
\noindent\textbf{Learning process.}
%We display the learning curve of the aggregation weights and the global validation accuracy of our "Auto-FedAvg-N-Dirichlet" model compared to "FedAvg-weighted" in Fig.~\ref{Fig:learn}. 
Here, we aim to analyze the learning process of Auto-FedAvg. The learning curve of the aggregation weights $\alpha$, validation accuracy growth, and the visualization of the Dirichlet distribution are displayed in Fig.~\ref{Fig:learn}. The sub-figures correspond to our best performing model ``Auto-FedAvg-N-Dichlet" in Table~\ref{tab:covid}. As shown in Fig.~\ref{fig:3a}, in the first 30 rounds, $\alpha_2$ and $\alpha_3$ rise moderately, indicating the global model could benefit from increasing the weight of the models from client \RM{2} and client \RM{3} in the early stage. This matches our expectation that client \RM{2} and client \RM{3} converge faster than client \RM{1} because client \RM{2} and client \RM{3} own significantly less data than client \RM{1}. Giving them more weight in the aggregation process accelerates the training process. As shown in Fig~\ref{fig:3b}, our approach has a faster growth in validation score than FedAvg. After approximately 40 rounds, we observe a rise of $\alpha_1$ and drops of $\alpha_2$ and $\alpha_3$, indicating that assigning higher weights to client \RM{1} benefits the global model eventually, making it more generalizable across different clients. 

In terms of the latent Dirichlet distribution of $\alpha$ (shown in Fig~\ref{fig:3c}), we plot the different states of $\alpha$ as well as the latent variable $\beta$ in round 0, 30, 50, and 300. Interestingly, the distribution becomes more concentrated with a smaller variance in round 300 compared to that of round 0. We interpret it as a higher certainty of the aggregation weights in the end of the training process than that in the beginning (starting from an initialization with $\beta=(6.0,6.0,6.0)$).

%\vspace{0.1cm}
%\noindent\textbf{Use training set to train the aggregation weights.} In our approach, we propose to learn the aggregation weights on the validation set. We also tried to update the aggregation weights on the training set with the other configurations staying the same as for ``Auto-FedAvg-N-Dirichlet". The performance on the test set turns out to be 61.53\%, which is 1.94\% lower than ``Auto-FedAvg-N-Dirichlet" using the validation set to learn the aggregation weights. We speculate that this is because each local model tends to overfit to its own local training set and strongly favours their own model when updating the aggregation weights $\alpha$. One evidence of this speculation is that the final aggregation weights $\alpha$ are approximately (0.3, 0.5, 0.2), indicating the local model of client \RM{2} is more favorable. In fact, client \RM{2} owns the smallest training set in which the segmentation model is the easiest to overfit. In contrast, if we use the validation, the generalizability of the local models can be measured more reliably, and thus produces more effective aggregation weights.

\vspace{0.1cm}
\noindent\textbf{The effect of interval of the aggregation weight learning.}

\begin{wrapfigure}[11]{r}{0.6\columnwidth}
%\hspace{-1cm}
\vspace{-0.45cm}
\includegraphics[width=0.6\columnwidth]{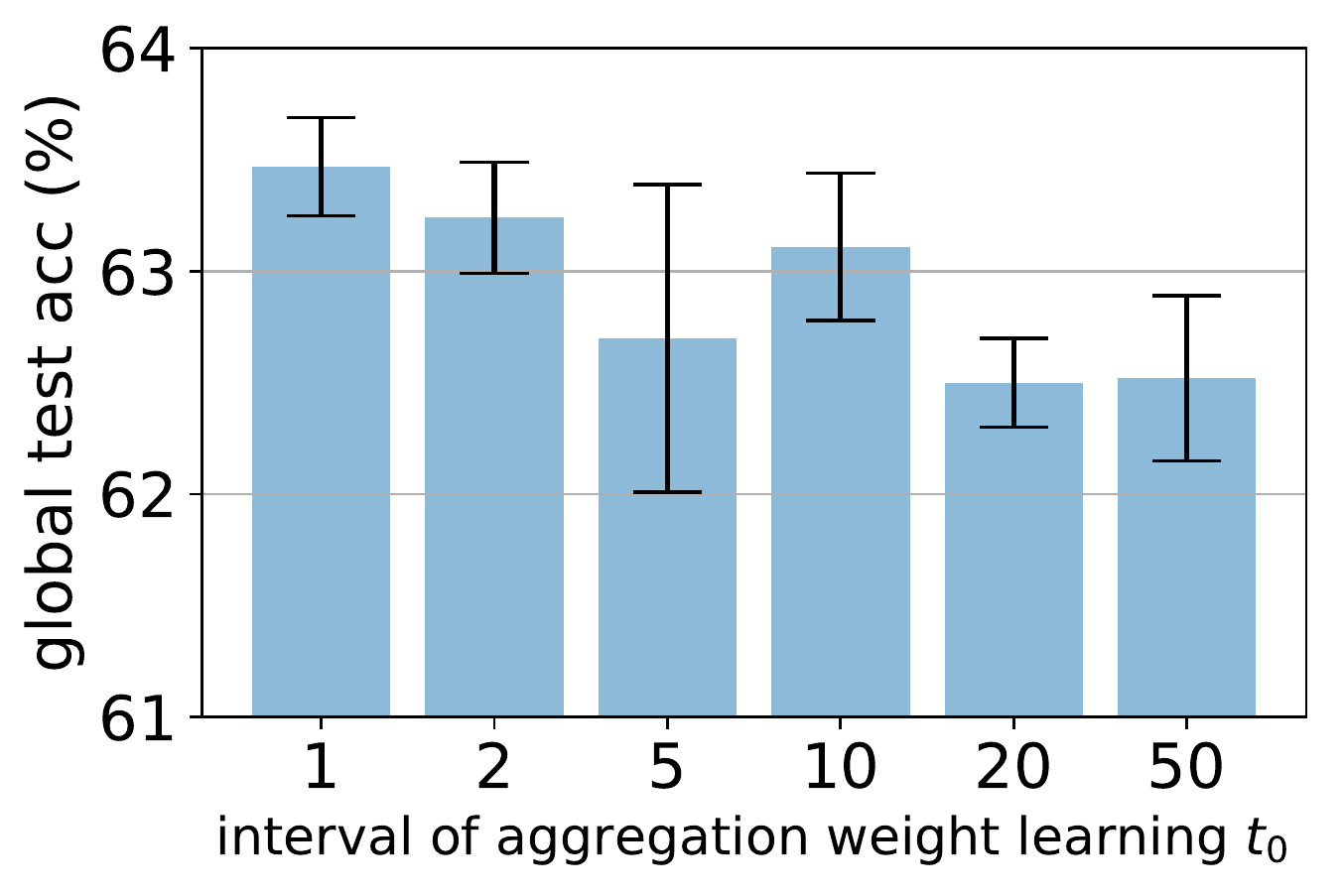}
%\vspace{-0.5cm}
\caption{The impact of the aggregation weight learning interval $t_0$.}
%\vspace{-0.5cm}
\label{fig:interval}
\end{wrapfigure}

We plot the global test accuracy under different aggregation weight learning intervals $t_0$ in Fig.~\ref{fig:interval}.

\setlength{\columnsep}{10pt}

The overall tendency is that the performance will degrade if we increase the interval $t_0$. It is worth mentioning that an interval of $t_0=10$ will only add to 10\% extra communication cost with merely a 0.36\% decline in performance compared to $t_0=1$. In terms of future applications of our approach, we expect no concern with such additional communication burden as medical institutions usually have relatively reliable network connectivity.

\vspace{0.1cm}
\noindent\textbf{Re-initialize aggregation weights before learning each round.}
Before each round of learning aggregation weights, we reuse the previously learned $\alpha$ from the last round as initialization. Another variant could be re-initializing $\alpha$ to be (0.33,0.33,0.33) before each round of aggregation weights learning (see \texttt{LearnAggWeight}, Alg. \ref{algo}). We conduct an experiment under this setting named ``Auto-FedAvg-N-Dirichlet*" in Table~\ref{tab:covid} and find the performance of test accuracy drops to $61.84\%$ compared to the original $63.11\%$. We speculate that this is because reusing the previous $\alpha$ not only offers a good starting point of the current round but also accelerates the learning process.

\subsection{Multi-institutional Pancreas Segmentation}
\noindent\textbf{Dataset description.} In this experiment, we study pancreas segmentation from CT scans, which is an important pre-requisite of pancreatic tumor detection and surgical planning~\cite{lowe2019precision}. We use the provided annotations from three public datasets, \emph{i.e.}, (i) pancreas segmentation subset of Medical Segmentation Decathlon~\cite{msd} which contains 281 cases (denoted as Dataset \RM{1}), (ii) the Cancer Image Archive (TCIA) Pancreas-CT dataset~\cite{tcia} which contains 82 cases (denoted as Dataset \RM{2}), and (iii) Beyond the Cranial Vault (BTCV) Abdomen data set~\cite{btcv} which contains 30 cases (denoted as Dataset \RM{3}). All the data include manual per voxel annotations of the pancreas from radiologists. For each dataset, we randomly split the annotated cases into training/validation/test sets, which are 95/93/93 for Dataset \RM{1}, 28/27/27 for Dataset \RM{2}, and 10/10/10 for Dataset \RM{3}. 

\vspace{0.1cm}
\noindent\textbf{Implementation details.} 
%As most of the implementation details are the same as for COVID-19 lesion segmentation, we only mention the differences here. 
In all FL experiments, we fix the number of total communication rounds to 50. In the local training process of each client, we adopt an initial learning rate of 0.001 with a cosine learning rate decay and with a batch size of 16. Same as the Covid-19 study, these hyperparameters are tuned to achieve the best local performances. 
%We resample each CT volume to a fixed spacing of (1.0mm, 1.0mm, 1.0mm) and randomly crop region of interests (ROIs) of $192^3$ with a foreground-background ratio as 1:1 to train the U-Nets. In the testing phase, we adopt a sliding window scheme with a stride of (48,48,48) and evaluate the performance on the original voxel spacing.

\vspace{0.1cm}
\noindent\textbf{Results.} We keep the same notation of our experiments as in Sec.~\ref{Sec:covid}. We found the conclusions are the same: our Auto-FedAvg outperforms FedAvg in all metrics and ``Auto-FedAvg-N-Dirichlet" is the best in both local performance and generalizability, indicating that the network-wise aggregation and using the Dirichlet distribution to model aggregation weights produce the best results. The conclusion is the same as of COVID dataset that network-wise formulation is better than layer-wise formulation and Dirichlet models aggregation weights better than the softmax. As for the significance test (Wilcoxon signed rank test) of the global model, we achieve significant improvements for two of three clients (datasets \RM{2} and \RM{3}), while the performance stays comparable to FedAvg for \RM{1} (with no significant difference). Interestingly, we find that with interval $t_0=5$, as denoted as ``Auto-FedAvg-N-Dirichlet*", the performance is even better than its $t_0=1$ counterpart (Auto-FedAvg-N-Dirichlet). This could result from the benefit of stabilization when the server keeps the aggregation weights fixed during the interval.

\section{Conclusions, Limitations, and Future Work}
In this paper, we introduced Auto-FedAvg, which improves the standard federated learning (FL) algorithm, FedAvg, by automatically and dynamically learning the aggregation weights instead of keeping them fixed. We also proposed a communication-efficient algorithm that alternates updates between the local model weights and the global aggregation weights. We further explored different constraints over the aggregation weights and variants of aggregation strategies. Experiments on the Cifar-10 and two extensive studies on real-world medical image analysis datasets illustrate the effectiveness of our approach.
%Experiments on two multi-institutional medical image segmentation datasets, \emph{i.e.}, COVID-19 lesion segmentation and pancreas segmentation, illustrated the effectiveness of our approach on real-world data. We also outperformed other state-of-the-art FL algorithms on a heterogeneous partitioning of the CIFAR-10 dataset \cite{fedma}.

One limitation of our algorithm is that relatively stable connections between the server and each of the clients are necessary. This is feasible in our ``cross-silo" situation but could be problematic in ``cross-device" scenarios where new edge devices regularly drop in or out~\cite{kairouz2019advances}. As a result, decreasing the communication frequency and integrating mechanisms for tolerating regular disconnections are two directions to improve the scalability of the current design. Our algorithm also introduced a general and flexible means to boost the performance of FL by updating a small number of global parameters and could be combined with differential privacy techniques for added protection against potential inversion attacks~\cite{fedmed1,kaissis2020secure}. 
Here, we explored only the network-wise and layer-wise learning of aggregation weights. Future research can include more complex 
aggregation operations and additional parameters to allow further
personalization for accounting for non-i.i.d data and domain shift cases in FL.
% However, more options are worth exploring, such as more complex aggregation operations and additional parameters to allow further personalization for addressing non-i.i.d issues and domain shifts in FL.

{\small
\bibliographystyle{unsrt}
\bibliography{egbib}
}

\end{document}